**Spectral emission dependence of tin-vacancy centers in diamond from thermal processing and chemical functionalization**

*Emilio Corte, Selene Sachero, Sviatoslav Ditalia Tchernij\*, Tobias Lühmann, Sébastien Pezzagna, Paolo Traina, Ivo Pietro Degiovanni, Ekaterina Moreva, Paolo Olivero, Jan Meijer, Marco Genovese, Jacopo Forneris*

E. Corte, Dr. S. Ditalia Tchernij, Prof. P. Olivero, Prof. J. Forneris
Dipartimento di Fisica, Università degli Studi di Torino
Istituto Nazionale di Fisica Nucleare (INFN), Sez. Torino
Istituto Nazionale di Ricerca Metrologica (INRiM)
via P. Giuria 1, 10125 Torino, Italy
E-mail: jacopo.forneris@unito.it

S. Sachero
Dipartimento di Fisica, Università degli Studi di Torino
via P. Giuria 1, 10125 Torino, Italy

Dr. T. Lühmann, S. Pezzagna, J. Meijer
Applied Quantum Systems, Felix-Bloch Institute for Solid-State Physics, Universität Leipzig
Linnéstraße 5, 04103 Leipzig, Germany

Dr. P. Traina, Dr. I.P. Degiovanni, Dr. E. Moreva, Dr. M. Genovese
Istituto Nazionale di Ricerca Metrologica (INRiM)
Strada delle Cacce 91, 10135 Torino, Italy

We report a systematic photoluminescence (PL) investigation of the spectral emission properties of individual optical defects fabricated in diamond upon ion implantation and annealing. Three spectral lines at 620 nm, 631 nm, and 647 nm are identified and attributed to the SnV center due to their occurrence in the PL spectra of the very same single-photon emitting defects. We show that the relative occurrence of the three spectral features can be modified by oxidizing the sample surface following thermal annealing. We finally report the relevant emission properties of each class of individual emitters, including the excited state emission lifetime and the emission intensity saturation parameters.

# 1. Introduction

Diamond-based color centers are appealing candidates as solid-state single-photon sources for applications in emerging fields of quantum technologies, such as quantum computing, quantum information and quantum sensing.[1-5] A class of group-IV-related quantum emitters, fabricated upon Sn ion implantation, thermal annealing and surface oxidation was recently discovered[6,7]. These color centers exhibit intriguing opto-physical properties, such as high emission rate, narrow spectral width, and emission concentrated in the zero-phonon line (ZPL), thus sparking the interest of the scientific community towards practical applications in the field of quantum technologies,[4] including quantum sensing,[8] nanophotonics,[9] and spintronics.[10,11]

The characteristic spectral feature of the optical activity of the SnV center consists of a strong room-temperature ZPL emission at 620 nm, which is widely attributed to the negative charge state of the defect (SnV⁻) based on the convincing support of independent theoretical works based on *ab initio* simulations.[6,11,12] In addition, multiple works have independently observed additional emission lines at 593 nm,[6,7,13,14] 631 nm,[7,14,15] 647 nm[6-8,13-15] and 663 nm.[12,14,15] While the 663 nm line has been convincingly interpreted as a radiation-induced defect,[12,15] the attribution of the remaining photoluminescence (PL) peaks is still uncertain and requires further disambiguation.

DFT simulations have highlighted the possibility to observe the SnV center in different charge states,[11] however a conclusive model of the experimental findings according to this interpretation has not been achieved yet. Recent reports on this subject suggest that the afore-mentioned emission lines are originating from different Sn-containing lattice complexes [6,8,9]. In these works the defects are created by ion irradiation, and different experimental conditions (ion implantation energy, annealing temperature and duration, and chemical termination of the latter) are explored. In particular, it is worth noting that the ion irradiation parameters result in different depth distributions of the defects, as well as their distance from the surface. Remarkably, high-pressure/high-temperature annealing processing performed without specific additional surface treatments resulted in the observation of the sole 620 nm emission line[6]. Moreover, the excitability of the 647 nm line under 633 nm laser excitation provided a strong experimental indication that such spectral feature cannot be attributed to the negative charge state of the SnV defect (620 nm ZPL)[15].

Finally, the electrical control of the Fermi level of an O-terminated p-i-p diamond junction has shown that all the 620 nm, 631 nm, 647 nm could be identified at the single-photon emitter level, but those spectral features were observed only from separate defects.[14]

In this work we compare the room-temperature properties of single emitters in a Sn-ion-implanted diamond sample following different post-implantation processing treatments. Particularly, we report the concurrent observation di of the 620 nm, 631 nm, and 647 nm from the same individual emitter after a thermal annealing process that was not followed by a chemical functionalization of the diamond surface. We also discuss the variation in the relative population of the emission lines in a discrete set of single centers upon the subsequent O-termination of the sample surface by means of plasma treatments. These results do not rule out the possible attribution of the 647 nm line to the optical activity of a different charge state of the SnV center. Its potential attribution to the SnV0 charge state[16], combining a long-lived spin coherence time, a brighter intensity with respect to the 620 nm line and an optical accessibility in the visible range for efficient and fast photon detection,[17] could pave the way towards the practical implementation of solid-state quantum memories.

## 2. Experimental

The measurements were performed on a high-purity single-crystal IIa diamond substrate, denoted as "electronic grade" due to the < 5 ppb concentration of both substitutional B and N impurities. A 200 × 200 μm² region was implanted using a collimated 60 keV Sn⁻ ions at 2 × 10¹⁰ cm⁻² fluence. The implantation was performed through a collimating steel mask placed at a ~5 mm distance from the sample surface. This configuration allowed the implantation of stray ions at a lower fluence at the outer edge of the implanted region. This outer region was selected to observe and investigate Sn-related emitters at the individual level. The ions penetration range - estimated as 23 nm according to SRIM (Stopping Range of Ions in Matter) simulations[18] - was sufficiently close to the sample surface to consider the chemical terminations to be effective at tuning possible charge states of the lattice defects.[19]

The sample was subsequently annealed for 2 h in vacuum at 950 °C to promote the formation of optically active defects. The photoluminescence (PL) characterization of individual defects was performed under two different functionalization conditions of the sample surface: "as implanted", i.e. without any subsequent functionalization following the thermal annealing process, and after a plasma treatment in $O_2$ (experimental parameters: 60 Pa pressure, 0.5 sccm $O_2$ flux, 30 min duration, 23 W microwave power).

The PL emission of individual SnV centers was characterized by means of a fiber-coupled single-photon sensitive room-temperature confocal microscope already presented in previous works.[20] The confocal microscope was operated using a 520 nm CW laser excitation. A 600 nm long-pass filter enabled the minimization of the background associated with the Raman scattering (567 nm); at the same time, it prevented the investigation of the 593 nm emission line reported in previous works.[6,7,13,14] The spectral features of the emitters were analysed using a single-grating monochromator (1200 grooves mm$^{-1}$, 600 nm blaze, ~4 nm spectral resolution) fiber-coupled to a single-photon avalanche detector.[21]

## 3. Results
### 3.1 Sn-related emitters formed upon thermal annealing

We report in this section the characterization of the emitters upon the sole thermal annealing, i.e. without any further surface functionalization. **Fig. 1a** shows a typical PL map (13x13 µm$^2$ area) acquired from the region under investigation[22], showing a density of ≤1 µm$^{-2}$ bright luminescent spots, whose emission was investigated individually. The discussion in the following is focussed on those sole spots (~30), where HBT interferometry confirmed a second-order auto-correlation function $g^{(2)}(t=0)<0.5$ after background correction.[7] Consistently with what observed in the reports discussed in Sect. 1, three main groups of spectral lines were observed, corresponding to the emission at 620 nm (S1), 631 nm (S2) and 647 nm (S3). Selected spectra are reported in **Fig. 1b**, highlighting that the S1-S3 emission lines could be observed both separately (i-iii) as the sole spectral feature of the considered center, as well as concurrently in the emission spectra of the same individual single-photon emitter. Particularly, **Fig. 1b**-iv, v, and vi show the spectra of single-photon emitters exhibiting both the "S1 and S2", "S1 and S3", "S2 and S3" spectral features at the same time, respectively. **Fig. 2a** (grey dots refer to the as-annealed sample) shows the spectral position of the emission lines acquired during the PL characterization, evidencing a non-negligible (i.e. ~25%) occurrence of multiple emission lines among the S1-S3 set from the same point defect.

To provide further insight into the above discussion, we report in **Fig. 3a** the PL spectrum of an individual emitter exhibiting all the S1, S2, and S3 emission peaks. **Fig. 3b** shows the $g^{(2)}(t)$ acquired from the center under 0.27 mW laser excitation. The antibunching at t=0 (background-corrected $g^{(2)}(0)=0.42$, considering a signal-to-background ratio[6,7] of $\rho=0.54$) indicates that the emission originates from an individual lattice defect.[23]

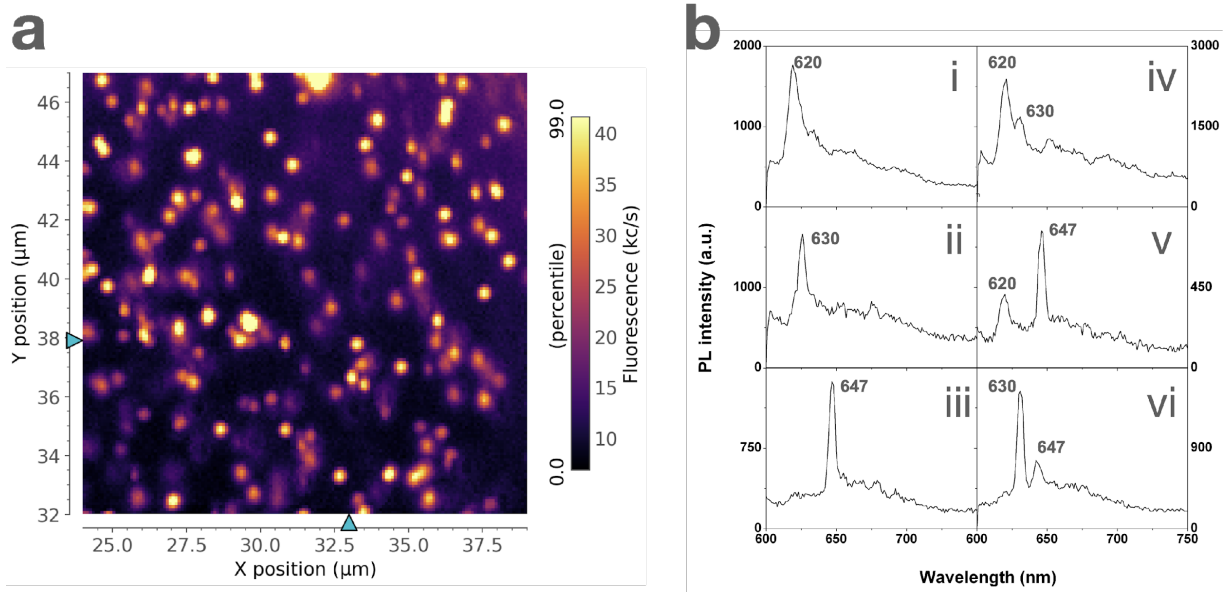

**Figure 1**. **a)** Typical confocal PL map acquired under 520 nm laser excitation from the Sn-implanted region of interest of the sample. **b)** PL spectra from individual emitters exhibiting emission lines at: (i) 620 nm, (ii) 631 nm, (iii) 647 nm, (iv) 620 nm and 631 nm; (v) 620 nm and 647 nm, (vi) 631 nm and 647 nm.

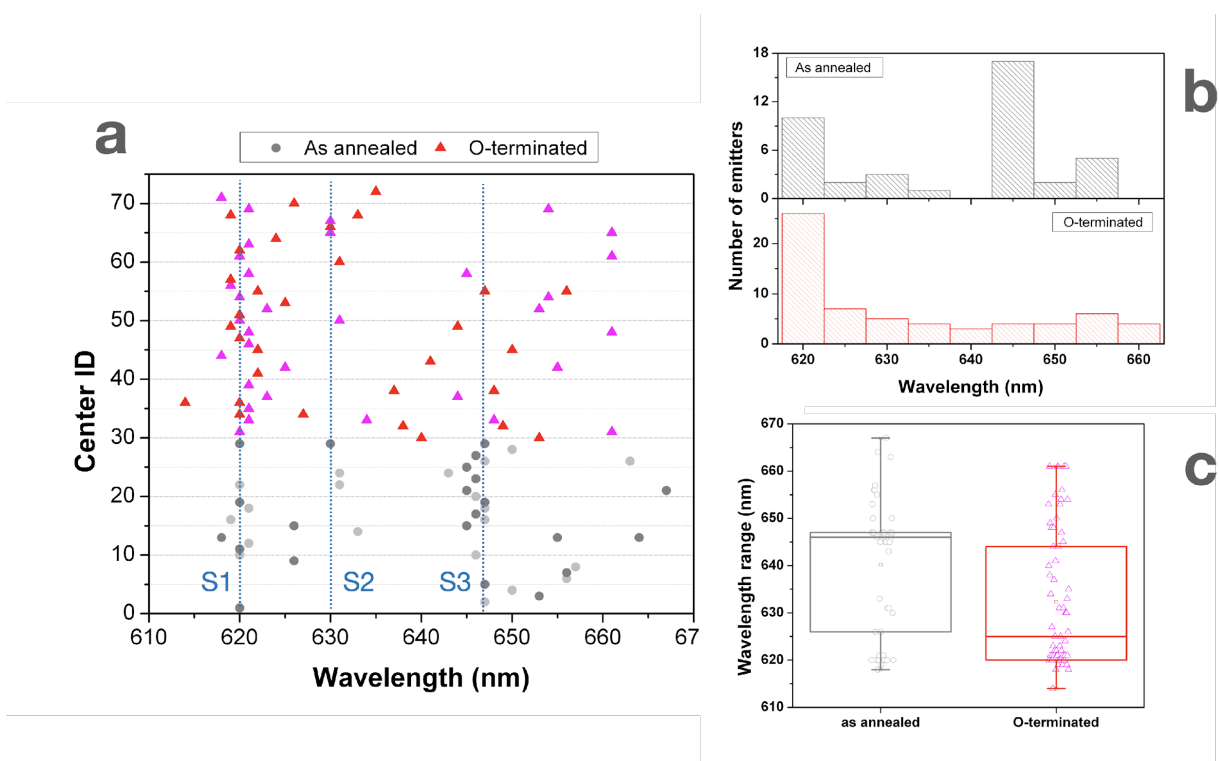

**Figure 2**. **a)** Typical confocal PL map acquired under 520 nm laser excitation from the Sn-implanted region of interest of the sample. **b)** PL spectra from individual emitters exhibiting emission lines at: (i) 620 nm, (ii) 631 nm, (iii) 647 nm, (iv) 620 nm and 631 nm; (v) 620 nm and 647 nm, (vi) 631 nm and 647 nm.

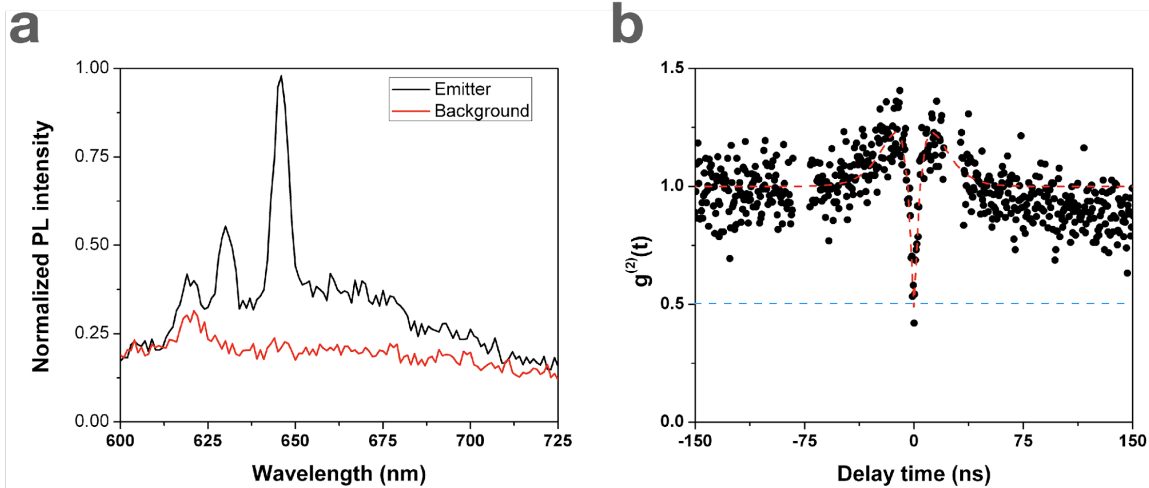

**Figure 3**. **a)** Emission spectrum from a single emitter (black line) showing the S1, S2, S3 lines. **b)** Background-corrected second-order autocorrelation function g$^{(2)}$(t) acquired under 520 nm laser excitation (0.27 mW power).

### 3.2 Sn-related emitters characterization upon surface O-termination

We report the PL characterization of the emitters in the region of interest of the sample following the $O_2$ plasma treatment described in Sect. 2. The O-termination of the sample required to unmount the sample from the confocal microscope. After this process the same region of the sample was investigated, but a study of the very same emitters studied in the previous experimental campaigns could not be performed. The interpretation of the data therefore rely on the assumption that the occurrence of the spectral lines is homogeneous in the whole region of interest, provided that the same density of centers per surface unit was observed. Additional characterizations performed after an $O_2$ plasma treatment performed at a higher microwave power (40 W) provided results statistically comparable with those reported here. **Fig. 2a** shows the spectral position of the PL lines observed over a set of ~40 emitters upon O-termination of the surface (red scatter data). The data indicate an apparent increase in the relative population of the S1 emission line, concurrently with a significant decrease in that of the S2 peak. This observation is further corroborated by the histogram in **Fig. 2b** showing the occurrence of the spectral lines as a function of the emission wavelength observed after thermal annealing (gray bars) and after plasma treatment (red).

The „as annealed" sample reveals a prevalence of the S3 line (~40% of the total amount of the observed lines) with respect to the S1 (25%). Such distribution is completely reversed in the O-terminated configuration, in which S1 becomes the most observed spectral feature (~40%), the S3 emission being associated to a mere 6% of the counts. Such observation is further evidenced by the box chart in **Fig. 2c**. Here, the median of the „as annealed" sample ZPL distribution (gray) coincides with the S3 emission line, while the range of spectra features compatible with the S1 line define the Q2-Q1 interquartile for the „O-terminated" data (red).

These data, together with the concurrent observation of both S1 and S3 lines from the very same individual defect, might be suggestive of their attribution to different charge states of the same defect. Particularly, the S1 line has been unambiguously attributed to the negative charge state of the tin-vacancy center SnV–.[6,11] The prevalence of this line upon O-termination of the diamond surface is also in line with what already reported for the nitrogen-vacancy center (NV), whose negative charge state is stabilized upon chemical functionalization by elements with positive electron affinity.[24] However, a direct comparison with the charge state population of NV centers could not be performed due to the scarce amount of these defects in the sample under investigation. Therefore, it could not be completely ruled out the role of concurring local strain, originating from an incomplete annealing of the radiation-induced defects introduced upon ion implantation, under the parameters adopted for the sample processing.[6]

The histogram in **Fig. 2b** does not reveal further significant spectral components, neither in the as-annealed sample, nor following the subsequent surface oxidization. Both the S2 and the 660 nm lines are observed, although their occurrence is significantly smaller than that of the main S1 and S2 spectral features (S2: 7% as-annealed, 8% O-terminated; 660 nm: 12% as annealed, 9% O-terminated).

It is also worth mentioning that 7% of the identified emission lines in the as-annealed sample does not correspond to the S1 or S3 lines, but occur in the spectral range comprised between 620 nm and 635 nm. Such fraction increases to 35% if the lines in the 625-655 nm range are considered for the O-terminated sample. This observation might not be incompatible with the hypothesis that the SnV center is not fully formed under the adopted low-pressure high-temperature annealing process[5,6], or with the interpretation of the ZPL spectral diffusion as a consequence of the radiation-induced local strain interacting with the defects.

### 3.3 Characterization of emitters with S1, S2, S3 spectral features

Individual Sn-related centers evidencing a single emission line were studied systematically in order to assess the respective emission properties in the "as annealed" sample. The properties observed at the single-photon emitter level did not reveal significant differences upon O-termination of the surface, besides the occurrence of each spectral line.

We report the results acquired from individual emitters exhibiting only one spectral feature, i.e. the S1, S2 and S3 line only, respectively.

Each of these defects was characterized by acquiring the second-order auto-correlation function curve at different optical excitation powers. Each of these curves underwent a fitting procedure according to a three-level system based on the following equation:[7,25]

$$g^{(2)}(t) = 1 - a_1 \cdot \exp(-|t| \cdot \lambda_1) + a_2 \cdot \exp(-|t| \cdot \lambda_2) \tag{1}$$

where the term proportional to $a_1$ describes the non classical anti-bunching signature of quantum emitters. The $\lambda_1$ parameter was fitted against the excitation power $P$ according to a linear model in order to infer the lifetime of the excited state as $\tau = [\lambda_1 (P=0)]^{-1}$.

We also studied the emission intensity $I$ of the defect as a function of the optical excitation power, in order to establish a preliminary comparison of the relevant emission properties of the three spectral components on the basis of those, already discussed in the literature, of the S1 line. The emission intensity was modelled after the expression[7,26]

$$I(P) = I_{sat} \cdot P / (P + P_{sat}) \tag{2}$$

Where $I_{sat}$ and $P_{sat}$ are the saturation intensity and the saturation excitation power, respectively.

*3.3.1 S1 - 620 nm emission line*

The characterization of the defect emitting at 620 nm ZPL (from whom the PL spectrum "i" reported in **Fig. 1b** was acquired) is shown in **Fig. 4**. $g^{(2)}(t)$ curves acquired in the 0.27-0.43 mW excitation power range (**Fig. 4a**) are reported as an example. The linear fitting of the $\lambda_1$ parameter (**Fig. 4b**) resulted in an estimation of the radiative emission lifetime of $\tau=(6.0\pm0.8)$ ns, value fully compatible with what reported in previous works and attributed to the ZPL of the SnV$^-$ center.[6,7,27]. The fitted emission intensity saturation curve (**Fig. 4c**) resulted in an excitation saturation power of $P_{sat} = (0.32\pm0.07)$ mW. This value is compatible with what reported in Ref. [7], considering that the $(1.11\pm0.01)$ mW value reported there quantified the laser power incident on the microscope objective. A decrease of ~60% in the power transmitted to the sample was observed in the confocal microscope adopted in this work. The saturation emission intensity $I_{sat}=(186\pm15)$ kcps was an order of magnitude lower than what reported in the previous characterization of this emission line.[7] This discrepancy is ascribed to the fiber-coupling of the source in the confocal microscope adopted for the present study, to be compared with the adoption of a higher-efficiency air-coupling of single-photon detectors in the referenced work. While a direct comparison between the two experiments cannot be performed without a metrological calibration of the two experimental apparatuses, the value obtained in this work could be used as a reference to assess the relative intensity of the S2, S3 emission lines with respect to the bright[6,7] emission of the 620 nm peak.

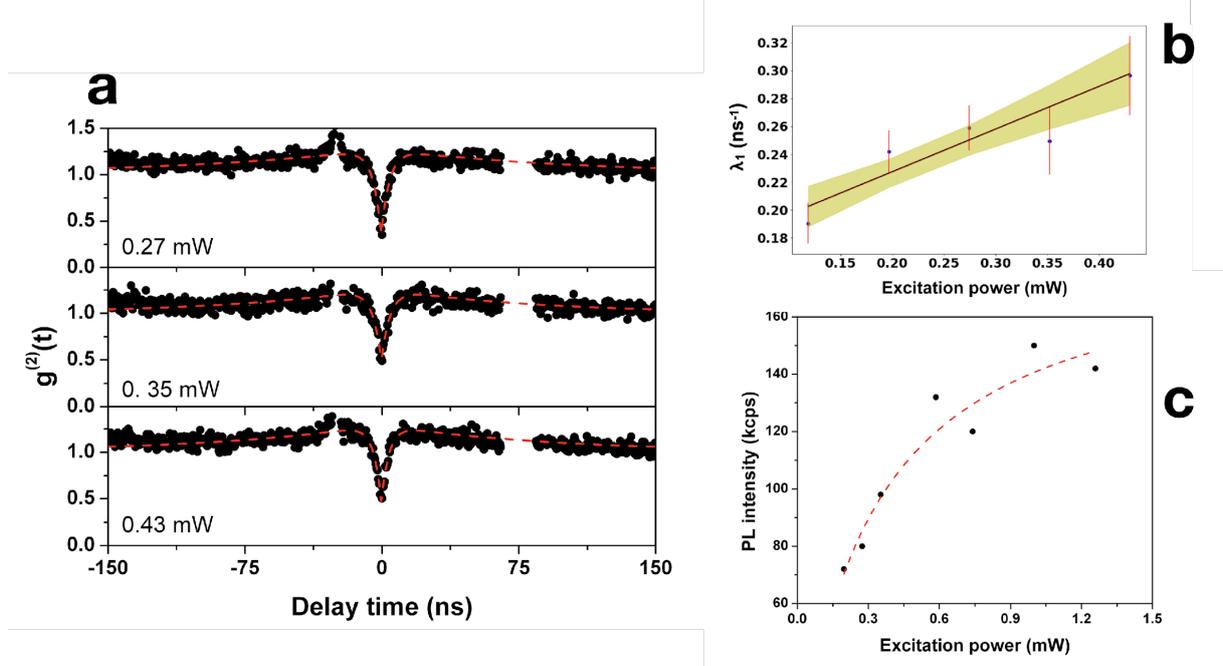

**Figure 4:** Characterization of the single-photon emission properties of an individual Sn-related defect emitting at the S1 line (620 nm, corresponding to the PL spectrum iii in Fig. 1b. **a)** second-order auto-correlation functions ($g^{(2)}(t)$ curves) acquired under 0.27 mW, 0.35 mW, 0.43 mW laser excitation power. **b)** Linear fit of the decay constant $\lambda_1$ extracted from the function fit of the $g^{(2)}(t)$ curves as a function of the excitation power. Red lines represent the fitting curves. **c)** Emission rate as a function of the laser excitation power.

*3.3.2 S2 - 631 nm emission line*
A defect emitting at 631 nm (PL spectrum "ii" in **Fig. 1b**) was characterized against the same emission properties (**Fig. 5**). The acquisition of $g^{(2)}(t)$ curves in the 0.27-0.59 mW excitation power range (**Fig. 5a**) enabled a preliminary estimation of the center lifetime as $\tau = \lambda_1(0)^{-1} = (4.8 \pm 1.7)$ ns (**Fig. 5b**). Despite the value being compatible with that of the S1 line, the saturation emission intensity (**Fig. 5c**) displayed a significantly higher value of $I_{sat}=(300\pm40)$ kcps achieved at a compatible excitation saturation power $P_{sat}=(0.42\pm0.08)$ mW. The compatibility of the relevant physical parameters such as $\tau$ and $P_{sat}$, together with the frequent observation of the peaks from the very same defect, could not enable us to completely rule out that the S1 and S2 lines are related to the same emission from the negatively-charged SnV defect, a hypothesis that would be in line with the tentative attribution of the S2 spectral feature to the first phonon replica of the S1 ZPL.[6]

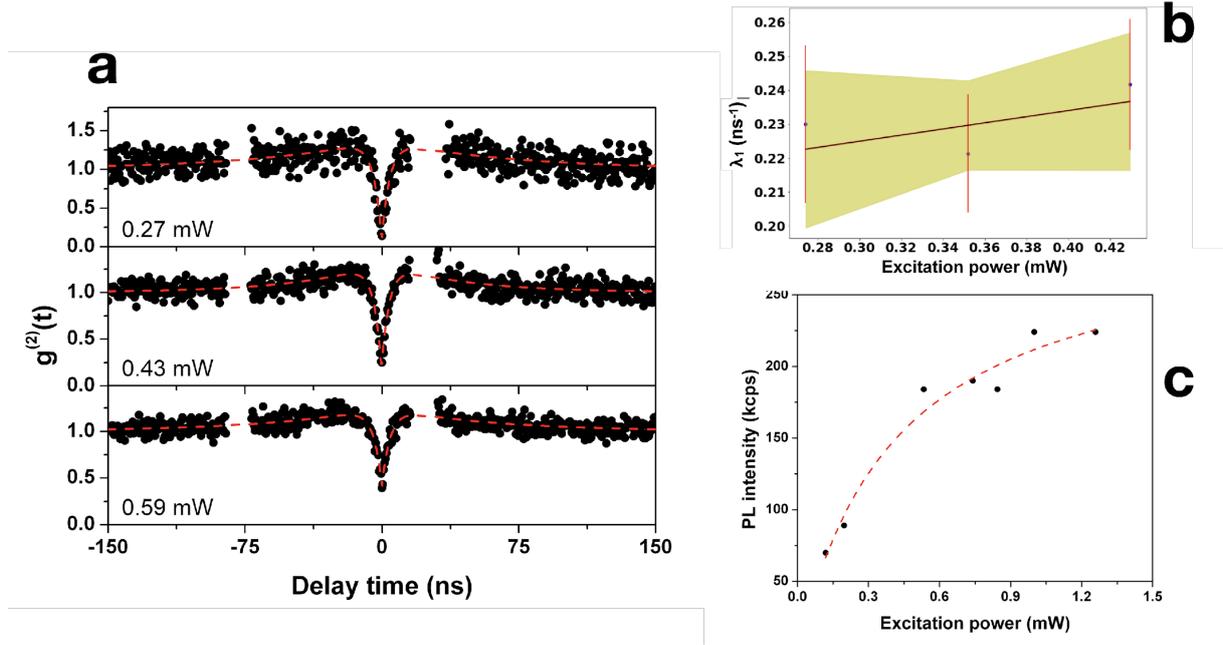

**Figure 5:** Characterization of the single-photon emission properties of an individual Sn-related defect emitting at the S2 line (631 nm, corresponding to the PL spectrum ii in Fig. 1b. **a)** second-order auto-correlation functions ($g^{(2)}(t)$ curves) acquired under 0.27 mW, 0.43 mW, 0.59 mW laser excitation power. **b)** Linear fit of the decay constant $\lambda_1$ extracted from the function fit of the $g^{(2)}(t)$ curves as a function of the excitation power. **c)** Emission rate as a function of the laser excitation power. Red lines represent the fitting curves.

*3.3.3 S3 - 647 nm emission line*

The $g^{(2)}(t)$ curves relative to the 647 nm emission of a defect associated with the PL spectrum "iii" in **Fig. 1b** are shown in (**Fig. 6a**) for the 0.27-0.59 mW excitation power range. The measurements enabled to quantify the center's excited state lifetime as $\tau=(7.4\pm1.0)$ ns (**Fig. 6b**), compatible with what was found for the S1 line but not for S2. The saturation emission intensity (**Fig. 5c**) displayed a value of $I_{sat}=(450\pm70)$ kcps, suggesting that the center is twice as bright as the SnV$^-$ ZPL. Conversely, the optical power required to achieve the saturation ($P_{sat}=(1.0\pm0.3)$ mW) is twice as higher.

A summary of the relevant characterization parameters is reported in **Table 1** for ease of comparison.

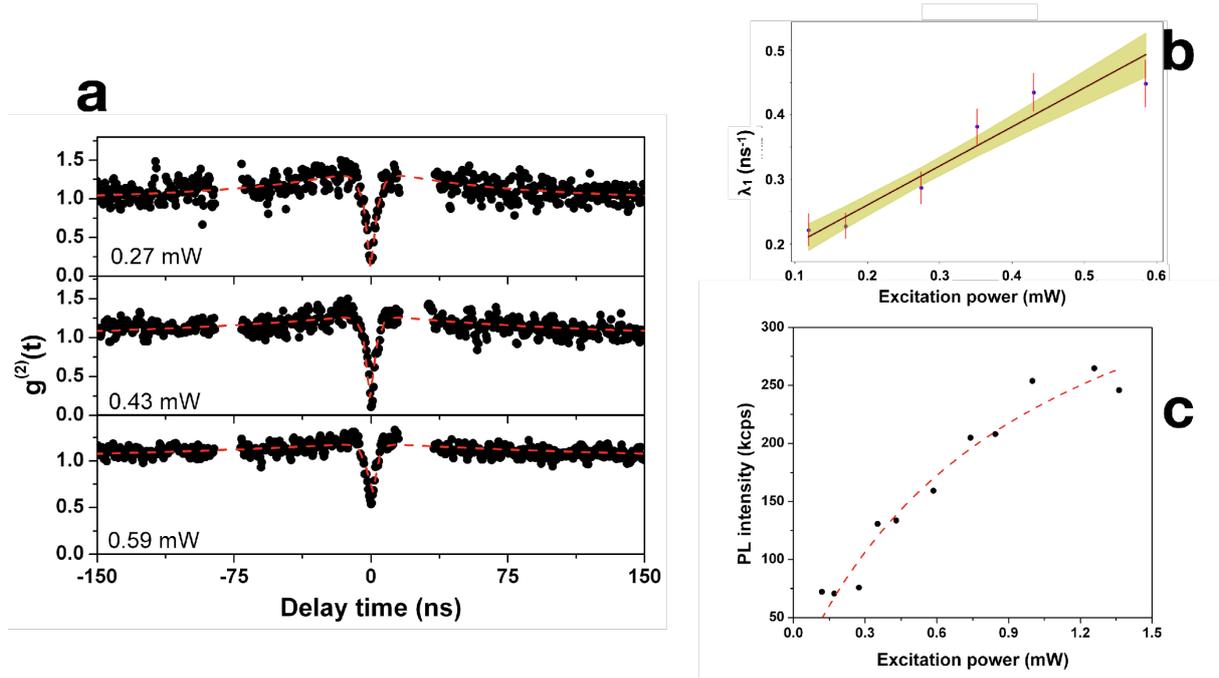

**Figure 6:** Characterization of the single-photon emission properties of an individual Sn-related defect emitting at the S3 line (647 nm, corresponding to the PL spectrum iii in Fig. 1b. **a)** second-order auto-correlation functions ($g^{(2)}(t)$ curves) acquired under 0.27 mW, 0.43 mW, 0.59 mW laser excitation power. **b)** Linear fit of the decay constant $\lambda_1$ extracted from the function fit of the $g^{(2)}(t)$ curves as a function of the excitation power. **c)** Emission rate as a function of the laser excitation power. Red lines represent the fitting curves.

| Wavelength (nm) | Emission saturation intensity (kcps) | Emission saturation power (mW) | Excited state lifetime (ns) |
|---|---|---|---|
| 620 | 186 ± 15 | 0.32 ± 0.07 | 6.0 ± 0.8 |
| 631 | 300 ± 40 | 0.42 ± 0.08 | 4.8 ± 1.7 |
| 647 | 450 ± 70 | 1.0 ± 0.3 | 7.4 ± 1.0 |

**Table 1.** Summary of the emission properties of individual defects in Sn-implanted diamond.

## Conclusions

In this work we reported the observation of multiple spectral emission lines at 620 nm (S1), 631 nm (S2) and 647 nm (S3) from individual defects formed upon Sn ion implantation in diamond and subsequent annealing, all of which were consistently observed in multiple previous reports.[6-8,12,14] The data presented here show for the first time the concurrent observation of all these three spectral contributions from the very same, single-photon emitting defect, suggesting their attribution to the same defect complex, i.e. the SnV center.

We also investigated the occurrence of the S1-S3 lines as a function of the sample processing. The as-annealed substrate exhibited a prevalence of the S3 spectral feature among the spectral emission of ~40 individual centers. Conversely, the S3 population density decreased in favour of the S1 emission line upon O-termination of the sample surface-. This observation, in combination with the observation of both lines from the very same defect and of different emission lifetimes and intensity excitation saturation parameters, is suggestive of the attribution of the 647 nm PL peak to a different charge state, possibly neutral charge state of the SnV complex (SnV$^0$), which has not been observed so far in other promising group-IV-related centers such as the GeV and the PbV.[4] This attribution will require a confirmation on centers fabricated upon chemical vapor deposition synthesis or ion implantation followed by HPHT annealing, in order to rule out the role of local strains in the emergence of the 647 nm line. If confirmed, in analogy with what reported for the SiV$^0$ center,[28] a neutrally charged SnV complex might display strong insensitivity to environmental noise and therefore offer further increased spin coherence times and higher brightness, thus proving to be a natural candidate for the implementation of long-lived quantum memories.

**Acknowledgements**

This work was supported by the following projects:

Coordinated Research Project "F11020" of the International Atomic Energy Agency (IAEA); "Piemonte Quantum Enabling Technologies" (PiQuET) project funded by the Piemonte Region within the "Infra-P" scheme (POR-FESR 2014-2020 program of the European Union); "Departments of Excellence" (L. 232/2016), funded by the Italian Ministry of Education, University and Research (MIUR); "Ex post funding of research" project of the University of Torino funded by the "Compagnia di San Paolo"; "Intelligent fabrication of QUANTum devices in DIAmond by Laser and Ion Irradiation" (QuantDia) project funded by the Italian Ministry for Instruction, University and Research within the "FISR 2019" program; "Training on LASer fabrication and ION implantation of DEFects as quantum emitters" (LasIonDef) project funded by the European Research Council under the "Marie Skłodowska-Curie Innovative Training Networks" program; CSN5 "PICS4ME" experiment funded by the Italian National Institute of Nuclear Physics (INFN); "Single-photon sources as new quantum standards" (SIQUST) project: 17FUN06 SIQUST has received funding from the EMPIR programme co-financed by the Participating States and from the European Union's Horizon 2020 research and innovation programmes; "Beyond Classical Optical Metrology" (BeCOMe) project: 17FUN01 (BeCOMe) leading to this publication has received funding from the EMPIR programme co-financed by the Participating States and from the European Union's Horizon 2020 research and innovation programme. T.L., S.P and J.M. acknowledge the support of the ASTERIQS program of the European Commission.